# Memristors - from In-memory computing, Deep Learning Acceleration, Spiking Neural Networks, to the Future of Neuromorphic and Bio-inspired Computing

*Adnan Mehonic [\*], Abu Sebastian, Bipin Rajendran, Osvaldo Simeone, Eleni Vasilaki, Anthony J. Kenyon*

Dr. Adnan Mehonic, Prof Anthony J. Kenyon
Department of Electronic & Electrical Engineering, UCL, Torrington Place, London WC1E 7JE, United Kingdom
E-mail: adnan.mehonic.09@ucl.ac.uk

Dr. Abu Sebastian
IBM Research – Zurich, 8803 Rüschlikon, Switzerland

Dr. Bipin Rajendran, Prof Osvaldo Simeone
Centre for Telecommunications Research, Department of Engineering, King's College London, WC2R 2LS, United Kingdom

Prof. Eleni Vasilaki
Department of Computer Science, University of Sheffield, Sheffield, South Yorkshire, United Kingdom

## Abstract

Machine learning, particularly in the form of deep learning, has driven most of the recent fundamental developments in artificial intelligence. Deep learning is based on computational models that are, to a certain extent, bio-inspired, as they rely on networks of connected simple computing units operating in parallel. Deep learning has been successfully applied in areas such as object/pattern recognition, speech and natural language processing, self-driving vehicles, intelligent self-diagnostics tools, autonomous robots, knowledgeable personal assistants, and monitoring. These successes have been mostly supported by three factors: availability of vast amounts of data, continuous growth in computing power, and algorithmic innovations. The approaching demise of Moore's law, and the consequent expected modest improvements in computing power that can be achieved by scaling, raise the question of whether the described progress will be slowed or halted due to hardware limitations. This paper reviews the case for a novel beyond-CMOS hardware technology – memristors – as a potential solution for the implementation of power-efficient in-memory computing, deep learning accelerators, and spiking neural networks. Central themes are the reliance on non-von-Neumann computing architectures and the need for developing tailored learning and inference algorithms. To argue that lessons from biology can be useful in providing directions for further progress in artificial intelligence, we briefly discuss an example based reservoir computing. We conclude the review by speculating on the "big picture" view of future neuromorphic and brain-inspired computing systems.

1. **Introduction**

The three factors are currently driving the main developments in artificial intelligence (AI): availability of vast amounts of data, continuous growth in computing power, and algorithmic innovations. Graphics processing units (GPUs) have been demonstrated as effective co-processors for the implementation of machine learning (ML) algorithms based on deep learning (DL). Solutions based on deep learning and GPU implementations have led to massive improvements in many AI tasks, but have also caused an exponential increase in demand for computing power. Recent analyses show that the demand for computing power has increased by a factor of 300,000 since 2012, and the estimate is that this demand will double every 3.4 months – at a much faster rate than improvements made historically through Moore's scaling (a 7-fold improvement over the same period of time) [1]. At the same time, Moore's law has been slowing down significantly for the last few years [2], as there are strong indications that we will not be able to continue scaling down CMOS transistors. This calls for the exploration of alternative technology roadmaps for the development of scalable and efficient AI solutions.

Transistor scaling is not the only way to improve computing performance. Architectural innovations such as GPUs, field-programmable arrays (FPGAs), and application-specific integrated circuits (ASICs), have all significantly advanced the ML field[3]. A common aspect of modern computing architectures for ML is a move away from the classical von Neumann architecture that physically separates memory and computing. This approach yields a performance bottleneck that is often the main reason for both energy and speed inefficiency of ML implementations on conventional hardware platforms due to costly data movements. However, architectural developments alone are not likely to be sufficient. In fact, standard digital CMOS components are inherently not well suited for the implementation of a massive number of continuous weights/synapses in artificial neural networks (ANNs).

**1.1. The promise of memristors.** There is a strong case to be made for the exploration of alternative technologies. Although the memristor technology is currently still in development, it is a strong candidate for future non-CMOS and beyond von-Neumann computing solutions [4]. Since its early development in 2008[5], or even earlier under different names [6], memristor technology expanded remarkably to include many different materials solutions, physical mechanisms, and novel computing approaches [4]. A single progress report cannot cover all different approaches and fast-growing developments in the field. The evaluation of state of the art in memristor-based electronics can be found elsewhere [7]. Instead, in this paper, we present and discuss a few representative case studies, showcasing the potential role of memristors in the expanding field of AI hardware. We present examples of how memristors are used for in-memory computing systems, deep learning accelerators, and spike-based computing. Finally, we discuss and speculate on the future of neuromorphic and bio-inspired computing paradigms and provide reservoir computing as an example.

For the last 15 years, memristors have been a focal point for many different research communities - mathematicians, solid-state physicists, experimental material scientists, electrical engineers and, more recently, computer scientists and computational neuroscientists. The concept of memristor was introduced almost 50 years ago, back in 1971[8], was nearly forgotten for almost four decades. It is now experiencing a rebirth with a vibrant and very active research community. There are many different flavours of memristive technologies. Still, in their most popular implementation, memristors are simple two-terminal devices with the extraordinary property that their resistance depends on their history of electrical stimuli. In other words, memristors are resistors with memory. They promise high levels of integration, stable non-volatile resistance states, fast resistance switching, excellent energy efficiency - all very desirable properties for next generation of memory technologies.

The physical implementations of memristors are broad and arguably include many different technologies such as redox-based resistive random-access memory (ReRAM), phase change memories (PCM), magnetoresistive random-access memory (MRAM). Further differentiations within larger classes can be made, depending on physical mechanisms that govern the resistance change. Many excellent reviews cover the principles and switching mechanisms of memristor devices. Here, we will briefly mention two extensively studied types of memristive devices, namely redox-based random access memory (ReRAM) and phase-change memory (PCM).

Resistance switching is one of the most explored properties of memristive devices. A thin insulating film reversibly changes its electrical resistance – between an insulating state and a conducting state – under the application of an external electrical stimulus. For binary memory devices, two stable states are sought, typically called the high resistance state (HRS), and the low resistance state (LRS). The transition from the HRS to the LRS is called a SET process, while a RESET process describes the transition from the LRS to the HRS.

Basic memory cells of both types, in their most straightforward implementation, have three layers – two conductive electrodes and a thin switching layer sandwiched in-between. Local redox processes govern resistance switching in ReRAM devices. A broad classification can be made based on a distinction between the switching that happens as a result of intrinsic properties of the switching material (typically oxides), and switching that is the result of in-diffusion of metal ions (typically from one of the metallic electrodes). The former type is called intrinsic switching, and the latter is called extrinsic switching[9]. Alternatively, a classification can be made depending on the main driving force for the redox process (thermal or electrical), or the type of ions that move. The main three classes are electrochemical metallization cells (or conductive bridge) ReRAMs (ECM), valence change ReRAMs (VCM) and thermochemical ReRAMs (TCM)[4].

Many ReRAM devices require an electroforming step prior to resistance switching. This can be considered a soft breakdown of the insulating material. A conductive filament is produced inside the insulating film as a result of the applied electrical bias. Modification of conductive filaments, led by a local redox process, leads to the change of resistance. The diameter of the conductive filament is typically of the order of a few nanometers to a few tens of nanometers, and it does not depend on the size of the electrodes. Another, less common type is interface-type switching, which does not depend on creation and modification of conductive filaments, but can be driven by the formation of a tunnel or Schottky barrier across the whole interface between electrode and switching layer.

In the case of PCMs, the change of resistance due to the crystallisation and amorphisation processes of phase change materials. Amplitude and duration of applied voltage pulses control the phase transitions – the SET process changes the amorphous to a crystalline phase (HRS to LRS transition), and the RESET process changes the crystalline to an amorphous phase (LRS to HRS transition).

For many computing tasks, more than two states are required, and for most memristive devices, including ReRAMs and PCMs, many resistance states can be achieved. However, benchmarking of memristive devices for different applications, beyond pure digital memory, can be challenging and relies on many different parameters other than the number of different resistance states. We will discuss the main device properties in the context of different applications.

**1.2 The landscape of different approaches and applications.** In the context of this paper, memristors can be used in applications beyond simple memory devices [10]. A "big picture" landscape of memristor-based approaches for AI is shown in Figure 1. There is more than one way that memristors can perform computing. A unique feature of memristor devices is the ability to co-locate memory and computing and to break the von Neumann bottleneck at the lowest, nanometre-scale level. One such approach is the concept of in-memory computing, which uses memory not only to store the data but also to perform computation at the same physical location. Furthermore, memristors have long been considered for deep learning acceleration. Specifically, memristive crossbar arrays physically represent weights in artificial neural networks as conductances at each crosspoint. When voltages are applied at one side of the crossbar and current sensed on the orthogonal terminals, the array provides vector-matrix multiplication in constant time step using Kirchhoff's and Ohm's laws. Vector-matrix multiplications dominate most DL algorithms – hundreds of thousands are often needed during training and inference. When weights are implemented as memristor conductances, there is no need for the extensive power-hungry data movement required by conventional digital systems based on the von Neumann architecture.

Other more bio-realistic concepts are also being explored. These include schemes relying on spike-based communication. The central premise of this approach can be summarised with the motto "computing *with* time, not *in* time". It has been shown that memristors can directly implement some functions of biological neurons and synapses, most importantly, synapse-like plasticity, and neuron-like integration and spiking. In these solutions, the information is encoded and transferred in the form of voltage or current spikes. Memristor resistances are used as proxies for synaptic strengths. More importantly, adjustment of the resistances is controlled according to local learning rules. One popular local learning rule is spike-timing-dependent plasticity (STDP), which adjust a local state variable such as conductance dynamically based on the relative timing of spikes. In a simple example, the conductance of a memristive "synapse" can be increased or decreased depending on the degree of overlap between pre- and post-synaptic voltage pulses. There also exist implementations that do not require overlapping pulses, instead utilising the volatile internal dynamics of memristive devices. Spike-based computing promises further improvements in power-efficiency, taking the inspiration from the remarkable efficiency of the human brain.

Finally, we speculate that, for future developments in AI, new knowledge and computational models from the fields of computational neuroscience could play a crucial role. Virtually all recent developments in ML and DL are driven by the field of computer science. At the same time, the algorithmic inspiration from neuroscience is mostly based on old models established as early as the 1950s. Although we are still at the infancy of understanding the full working principles of the biological brain, novel brain-inspired architectural principles, beyond simple probabilistic deep learning approaches, could lead to higher-level cognitive functionalities. One such example is the concept of reservoir computing, which we discuss briefly in the paper. It is unlikely that current digital CMOS transistor technology can be optimized for the implementation of much more dynamic and adaptive systems in an efficient way. In contrast, memristor-based systems, with their rich switching dynamics and many state variables, may provide a perfect substrate to build a new class of intelligent and efficient neuromorphic systems.

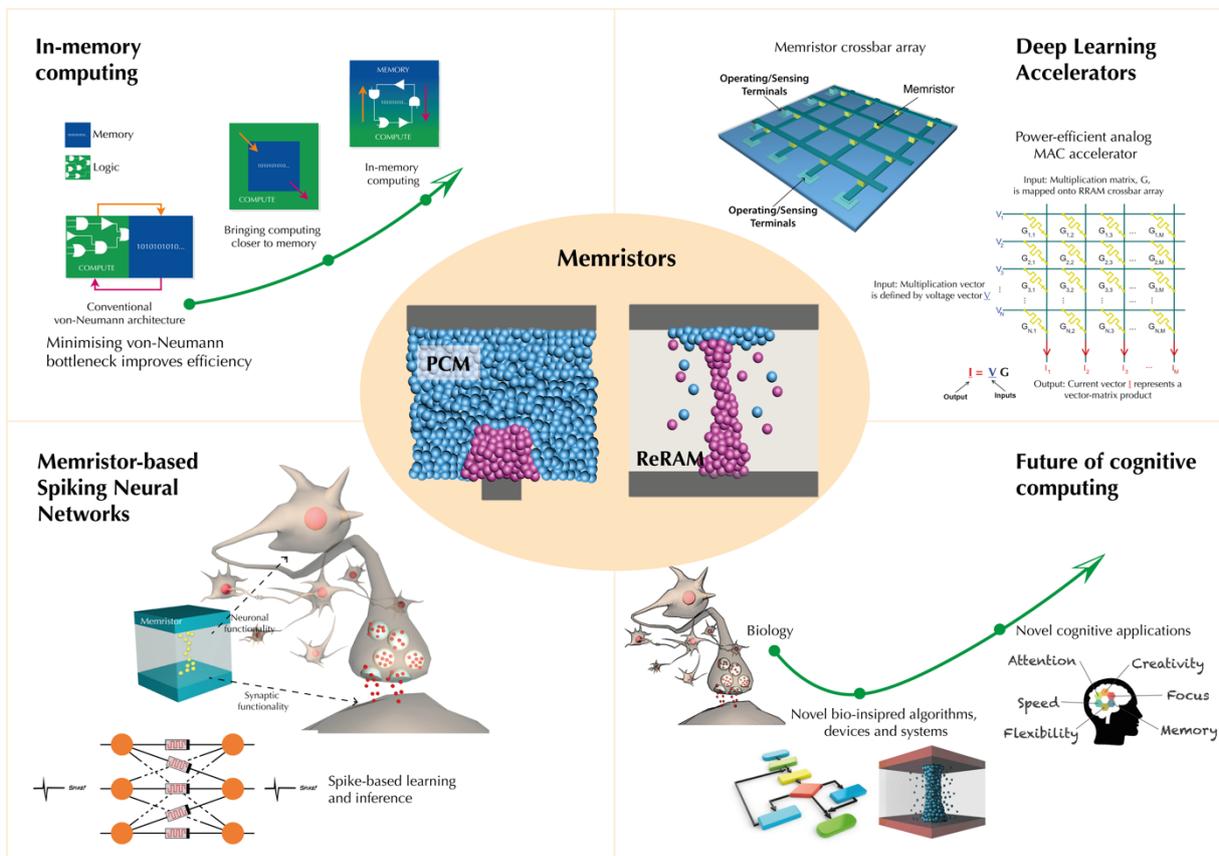

**Figure 1.** *The landscape of memristor-based systems for Artificial Intelligence. In-memory computing aims to eliminate the von-Neumann bottleneck by implementing compute directly within the memory. Deep learning accelerators based on memristive crossbars are used to implement vector-matrix multiplication directly using Ohm's and Kirchhoff's laws. Spiking neural networks, a type of artificial neural networks, are biologically more plausible and do not operate with continuous signals, but use spikes to process and transfer data. Memristor systems could provide a hardware platform to implement spike-based learning and inference. More complex functionalities (neuromorphic), beyond simple digital switching CMOS paradigm, directly implemented in memristive hardware primitives, might fuel the next wave of higher cognitive systems.*

## 2. In-memory computing

In the von Neumann architecture, which dates back to the 1940s, memory and processing units are physically separated and large amounts of data need to be shuttled back and forth between them during the execution of various computational tasks. The latency and energy associated with accessing data from the memory units are key performance bottlenecks for a range of applications, in particular for the increasingly prominent artificial intelligence related workloads [11]. The energy cost associated with moving data is a key challenge for both severely energy constrained mobile and edge computing as well as high performance computing in a cloud environment due to cooling constraints. The current approaches, such as using hundreds of processors in parallel [12] or application-specific processors [13], are not likely to fully overcome the challenge of data movement. It is getting increasingly clear that novel architectures need to be explored where memory and processing are better collocated. In-memory computing is one such non-von Neumann approach where certain computational tasks are performed in place in the memory itself organized as a computational memory unit [14,15,16,17]. As schematically illustrated in Figure 2, in-memory computing obviates the need to move data into a processing unit. Computing is performed by exploiting the physical attributes of the memory devices, their array-level organization, the peripheral circuitry as well as the control logic. In this paradigm, the memory is an active participant in the computational task. Besides reducing latency and energy cost associated with data movement, in-memory computing also has the potential to improve the computational time complexity associated with certain tasks due to the massive parallelism afforded by a dense array of millions of nanoscale memory devices serving as compute units. By introducing physical coupling between the memory devices, there is also a potential for further reduction in computational time complexity [18, 19]. Memristive devices such as PCM, ReRAM and MRAM [20, 21] are particularly well suited for in-memory computing.

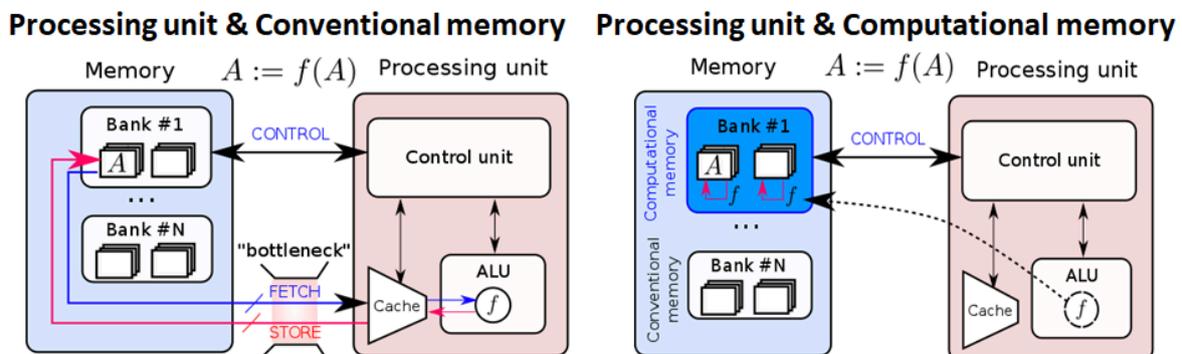

**Figure 2.** *In-memory computing. In a conventional computing system, when an operation f is performed on data D, D has to be moved into a processing unit. This incurs significant latency and energy cost and creates the well-known von Neumann bottleneck. With in-memory computing, f(D) is performed within a computational memory unit by exploiting the physical attributes of the memory devices. This obviates the need to move D to the processing unit. (Adapted and reproduced with permission [14], Copyright 2017, Nature Research)*

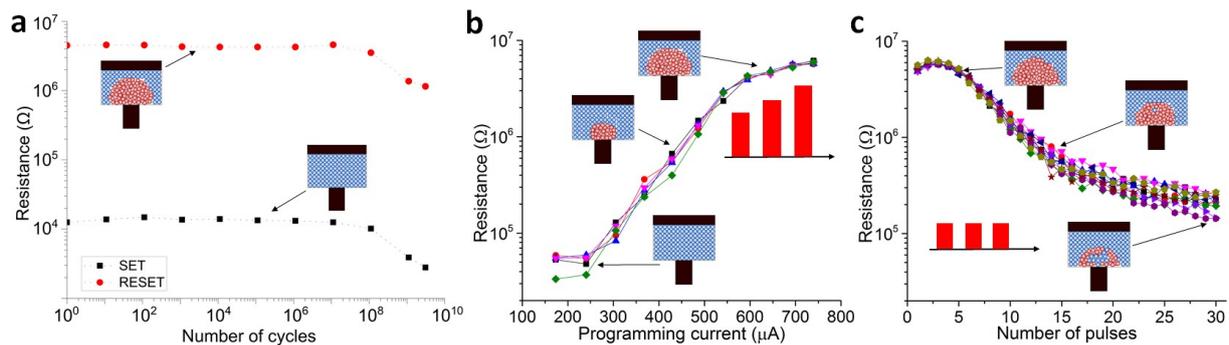

**Figure 3.** *The key physical attributes of memristive devices that facilitate in- memory computing. a) Binary storage capability whereby the devices can be switched between high and low resistance values in a repeatable manner (Adapted and reproduced with permission [22]. Copyright 2019, IOP Publishing). b) Multi- level storage capability whereby the devices can be programmed to a continuum of resistance values by the application of appropriate programming pulses (Adapted and reproduced with permission [23]. Copyright 2018, American Institute of Physics) c) The accumulative behavior whereby the resistance of a device can be progressively decreased by the successive application of identical programming pulses (Adapted and reproduced with permission [23]. Copyright 2018, American Institute of Physics).*

There are several key physical attributes that enable in-memory computing using memristive devices. First of all, the ability to store two levels of resistance/conductance values in a non-volatile manner and to reversibly switch from one level to the other (binary storage capability) can be exploited for computing. Figure 3**a** shows the resistance values achieved upon repeated switching of a representative PCM device between low resistance SET states and high resistance RESET states. Due to the SET and RESET states, resistance could serve as an additional logic state variable. In conventional CMOS, voltage serves as the single logic state variable. The input signals are processed as voltage signals and are output as voltage signals. By combining CMOS circuitry with memristive devices, it is possible to exploit the additional resistance state variable. For example, the RESET state could indicate logic '0' and the SET state could denote logic '1'. This enables logical operations that rely on the interaction between the voltage and resistance state variables and could enable the seamless integration of processing and storage. This is the essential idea behind memristive logic, which is an active area of research [24, 25, 26]. Memristive logic has the potential to impact application areas such as image processing [27], encryption and database query [28]. Brain-inspired hyperdimensional computing that involves the manipulation of large binary vectors has recently emerged as another promising application area for in-memory logic [29, 30]. Going beyond binary storage, certain memristive devices can also be programmed to a continuum of resistance or conductance values (analog storage capability). For example, Figure 3**b** shows a continuum of resistance levels in a PCM device achieved by the application of programming pulses with varying amplitude. The device is first programmed to the fully crystalline state, after which RESET pulses are applied with progressively increasing amplitude. The device resistance is measured after the application of each RESET pulse. Due to this property, it is possible to program a memristive device to a certain desired resistance value through iterative programming by applying several pulses in a closed-loop manner [31]. Yet another physical attribute that enables in-memory computing is the accumulative behavior exhibited by certain memristive devices. In these devices, it is possible to progressively reduce the device resistance by the successive application of SET pulses with the same amplitude. And in certain cases, it is possible to progressively increase the resistance by the successive application of RESET pulses. Experimental measurement of this accumulative behavior in a

PCM device is shown in Figure 3**c**. This accumulative behavior is central to applications such as training deep neural networks which is described later. The intrinsic stochasticity associated with the switching behavior in memristive devices can also be exploited for in-memory computing [32]. Applications include stochastic computing [33] and physically unclonable functions [34].

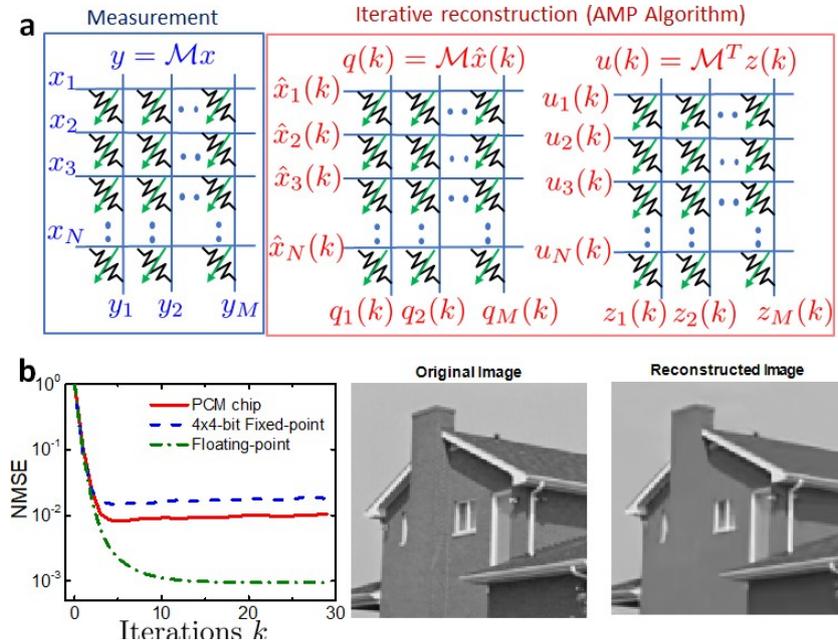

**Figure 4.** *a) Compressed sensing involves one matrix-vector multiplication. Data recovery is performed via an iterative scheme, using several matrix-vector multiplications on the very same measurement matrix and its transpose. b) An experimental illustration of compressed sensing recovery in the context of image compression is presented, showing 50% compression of a 128x128 pixel image. The normalized mean square error (NMSE) associated with the reconstructed signal is plotted against the number of iterations. Adapted and reproduced with permission* [35], *Copyright 2018, IEEE.*

A very useful in-memory computing primitive enabled by the binary and analog nonvolatile storage capability is matrix-vector multiplication (MVM) [36, 37]. The physical laws that are exploited to perform this operation are Ohm's law and Kirchhoff's current summation laws. For example, to perform the operation $Ax = b$, the elements of $A$ are mapped linearly to the conductance values of memristive devices organized in a crossbar configuration. The $x$ values are mapped linearly to the amplitudes of read voltages and are applied to the crossbar along the rows. The result of the computation, $b$, will be proportional to the resulting current measured along the columns of the array. Compressed sensing and recovery are one of the applications that could benefit from an in-memory computing unit that performs matrix-vector multiplications. The objective behind compressed sensing is to acquire a large signal at sub-Nyquist sampling rate and to subsequently reconstruct that signal accurately. Unlike most other compression schemes, sampling and compression are done simultaneously, with the signal getting compressed as it is sampled. Such techniques have widespread applications in the domain of medical imaging, security systems, and camera sensors. The compressed measurements can be thought of as a mapping of a signal $x$ of length $N$ to a measurement vector $y$ of length $M < N$. If this process is linear, then it can be modeled by an $M \times N$ measurement matrix $M$. The idea is to store this measurement matrix in the in-memory computing unit, with memristive devices organized in a cross-bar configuration (see Figure 4(a)). In this manner the compression operation can be performed in $O(1)$ time complexity.

To recover the original signal from the compressed measurements, an approximate message passing algorithm (AMP) can be used, using an iterative algorithm that involves several matrix-vector multiplications on the very same measurement matrix and its transpose. In this way the same matrix that was coded in the in-memory computing unit can also be used for the reconstruction, reducing reconstruction complexity from O(*MN*) to O(*N*). An experimental illustration of compressed sensing recovery in the context of image compression is shown in Figure 4(b). A 128x128-pixel image was compressed by 50% and recovered using the measurement matrix elements encoded in a PCM array. The normalized mean square error associated with the recovered signal is plotted as a function of the number of iterations. A remarkable property of AMP is that its convergence rate is independent of the precision of the matrix-vector multiplications. The lack of precision only results in a higher error floor, which may be considered acceptable for many applications. Note that, in this application, the measurement matrix remains fixed and hence the property of PCM that is exploited is the multi-level storage capability.

## 3. Deep learning accelerators

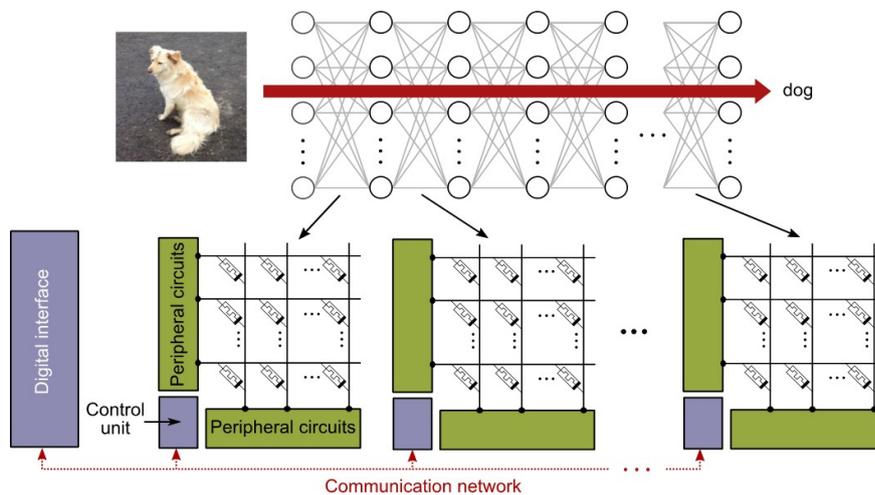

**Figure 5.** *Deep learning based on in-memory computing. The various layers of a neural network are mapped to a computational memory unit where memristive devices are organized in a crossbar configuration. The synaptic weights are stored in the conductance state of the memristive devices. A global communication network is used to send data from one array to another. Adapted and reproduced with permission* [17], *Copyright 2020, Nature Research.*

Deep neural networks (DNNs), loosely inspired by biological neural networks, consist of parallel processing units called neurons interconnected by plastic synapses. By tuning the weights of these interconnections using millions of labelled examples, these networks are able to perform certain supervised learning tasks remarkably well. These networks are typically trained via a supervised learning algorithm based on gradient descent. During the training phase, the input data is forward propagated through the neuron layers with the synaptic networks performing multiply-accumulate operations. The final layer responses are compared with input data labels and the errors are back-propagated. Both steps involve sequences of matrix-vector multiplications. Subsequently, the synaptic weights are updated to reduce the error. This optimization approach can take multiple days or weeks to train state-of-the-art networks on conventional computers. Hence, there is a significant effort towards the design of custom ASICs based on reduced precision arithmetic and highly optimized dataflow [13, 38]. However, the need to shuttle millions of synaptic weight values between the memory and processing unit remains a key performance bottleneck and hence in-memory computing is being explored as an alternative approach for both inference and training of DNNs [39, 40]. The essential idea is to map the various layers of a neural network to an in-memory computing unit where memristive devices are organized in a crossbar configuration (see Figure 5). The synaptic weights are stored in the conductance state of the memristive devices and the propagation of data through each layer is performed in a single step by inputting the data to the crossbar rows and deciphering the results at the columns.

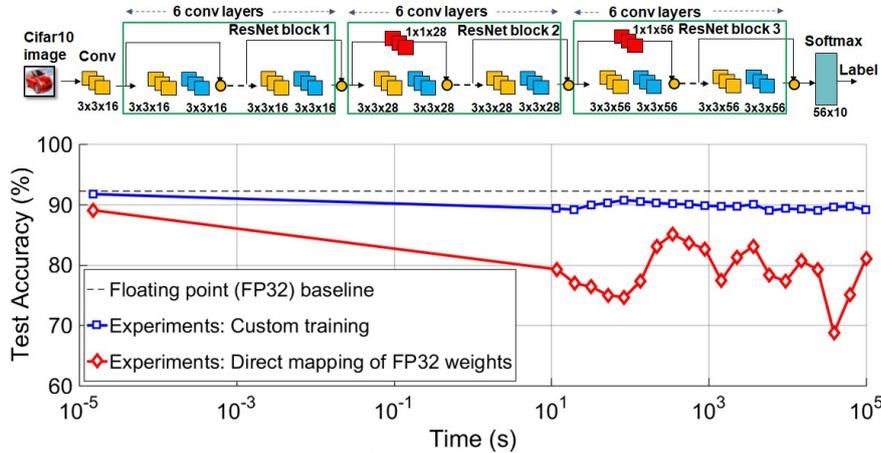

**Figure 6.** *Deep learning inference. Experimental results on ResNet-32 using the CIFAR-10 dataset. The classification accuracies obtained via the direct mapping and custom training approaches are compared to the floating-point baseline. Adapted and reproduced with permission [40], Copyright 2019, IEEE.*

Deep learning inference refers to just the forward propagation in a DNN once the weights have been learned. Both binary and analogue storage capability of memristive devices can be exploited for the MVM operations associated with the inference operation. The key challenges are the inaccuracies associated with programming the devices to a specified synaptic weight as well as drift, noise etc. associated with the conductance values[41]. Due to these reasons, the synaptic weights that are obtained by training in high precision arithmetic (e.g. 32-bit floating point) cannot be mapped directly to computational memory. However, it can be shown that by customizing the training procedure to make it aware of these device-level nonidealities, it is possible to obtain synaptic weights that are suitable for being mapped to an in-memory computing unit [42,40]. A more recent approach is to use the committee machines of multiple smaller neural networks. The approach shows the promise of increasing inference accuracy without increasing the number of devices by using a committee of smaller neural networks [43]. Figure 6 shows mixed hardware/software experimental results using a prototype multi-level PCM chip. The synaptic weights are mapped to PCM devices organized in a 2-PCM differential configuration (723,444 PCM devices in total). It can be seen that the custom training scheme approaches the floating-point base-line, whereas the direct mapping approach fails to deliver sufficient accuracy. The slight temporal decline in accuracy is attributed to the conductance drift exhibited by PCM devices [44]. However, in spite of the drift, a classification accuracy of close to 90% is maintained over a significant duration of time.

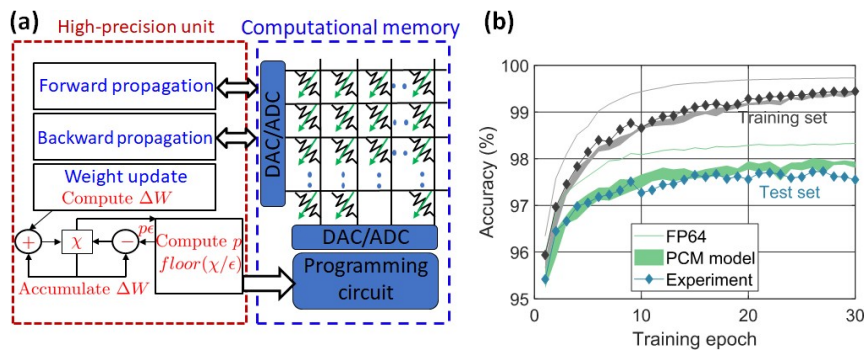

**Figure 7.** *Deep learning training. a) Schematic illustration of the mixed-precision architecture for training DNNs. b) The synaptic weight distributions and classification accuracies are compared between the experiments and floating point baseline[45].*

In-memory computing can also be used in the context of supervised training of DNNs with backpropagation. When performing training of a DNN encoded in crossbar arrays, forward propagation is performed in the same way as inference described above. Next, backward propagation is performed by inputting the error gradient from the subsequent layer onto the columns of the current layer and deciphering the result from the rows. Subsequently the error gradient is computed. Finally, the weight update is performed based on the outer product of activations and error gradients of each layer. This weight update relies on the accumulative behaviour of memristive devices. Recent deep learning research shows that when training DNNs, it is possible to perform the forward and backward propagations rather imprecisely while the gradients need to be accumulated in high precision [46]. This observation makes the DL training problem amenable to the mixed-precision in-memory computing approach that was recently proposed [47]. The in-memory compute unit is used to store the synaptic weights and to perform the forward and backward passes, while the weight changes are accumulated in high precision (Figure 7(a)) [48, 49]. When the accumulated weight exceeds a certain threshold, pulses are applied to the corresponding memory devices to alter the synaptic weights. This approach was tested using the handwritten digit classification problem based on the MNIST data set. A two-layered neural network was employed with 2-PCM devices in differential configuration (approx. 400,000 devices) representing the synaptic weights. Resulting test accuracy after 20 epochs of training was approx. 98% (Figure 7(b)). After training, inference on this network was performed for over a year with marginal reduction in the test accuracy. The crossbar topology also facilitates the estimation of the gradient and the in-place update of the resulting synaptic weight all in $O(1)$ time complexity [50, 39]. By obviating the need to perform gradient accumulation externally, this approach could yield better performance than the mixed-precision approach. However, significant improvements to the memristive technology, in particular the accumulative behavior, is needed to apply this to a wide range of DNNs [51, 52].

Compared to the charge-based memory devices that are also used for in-memory computing [53, 54, 55], a key advantage of memristive devices is the potential to be scaled to dimensions of a few nanometers [56, 57, 58, 59, 60]. Most of the memristive devices are also suitable for back end of line integration, thus enabling their integration with a wide range of front-end CMOS technologies. Another key advantage is the non-volatility of these devices that would obviate the need for computing systems to be constantly connected to a power supply. However, there are also challenges that need to be overcome. The significant intra-device and intra-device variability associated with the SET and RESET states is a key challenge for applications where memristive devices are used for logical operations. For applications that rely on analogue storage capability, a significant challenge is programming variability that captures the inaccuracies associated with programming an array of devices to desired conductance values. In ReRAM, this variability is attributed mostly to the stochastic nature of filamentary switching and one prominent approach to counter this is that of establishing preferential paths for CF formation [61, 62]. Representing single computational elements by using multiple memory devices is another promising approach [63]. Yet another challenge is the temporal and temperature-induced variations of the programmed conductance values. The resistance "drift" in PCM devices, which is attributed to the intrinsic structural relaxation of the amorphous phase, is an example. The concept of projected phase change memory is a promising approach towards tackling "drift" [64, 65]. The requirements that the memristive devices need to fulfil when employed for computational memory are heavily application dependant. For memristive logic, high cycling endurance ($> 10^{12}$ cycles) and low device-to-device variability of the SET/RESET resistance values are critical. For computational tasks involving read-only operations, such as matrix-vector multiplication, it is required that the conductance states remain relatively unchanged during their execution. It is also desirable to have a gradual analogue-type switching characteristic for programming a continuum of resistance values in a single device. A linear and symmetric accumulative behaviour is also required in applications where the device conductance needs to be incrementally updated such as in deep learning training [66]. For stochastic computing applications, random device variability is not problematic, but graceful device degradation is highly desirable, as described in [67].

## 4. Spiking Neural Networks and Memristors

As opposed to the deep learning networks discussed above, spiking neural networks (SNNs) can more naturally incorporate the notion of time in signal encoding and processing. SNNs are typically modelled on the integrate-and-fire behaviour of neurons in the brain. In this framework, neurons communicate with each other using binary signals or spikes. The arrival of a spike at a synapse triggers a current flow into the downstream neuron, with the magnitude of the current weighted by the effective conductance of the synapse. The incoming currents are integrated by the neuron to determine its membrane potential and a spike is issued when the potential exceeds a threshold. This spiking behaviour can be triggered in a deterministic or probabilistic manner. Once a spike is issued, the membrane potential is reset to a resting potential or decreased according to some predetermined rule. The integration is limited to a specific time window, or else a leak factor is incorporated in the integration, endowing the neuron model with a finite memory of past spiking events.

Compared to the realization of second-generation deep neural networks (DNNs discussed in the previous section), SNNs can potentially have significant improvements in efficiency. The first reason for this comes from the underlying signal encoding mechanism. The calculation of the output of a neuron involves the determination of the weighted sum of synaptic weights with real-valued neuronal outputs of the previous layer. For a fully connected second generation DNN with $N$ neurons in each layer, this requires $N^2$ multiplications of real valued numbers, typically stored in low precision representations. In contrast, the forward propagation operation in an SNN only requires addition operations, as the input neuronal signals are binary spike signals. To elaborate, assume that the input signal is encoded as a spike train with duration $T$, with a minimum inter-spike interval of $\Delta t$. If the probability of a spike at any instant of time is $p$, then on an average $NpT/\Delta t$ spikes have to be propagated through the synapses, and this requires $N^2pT/\Delta t$ addition operations. In most modern processors, the cost of multiplication, $C_m$, is 3-4 times higher than that of addition, $C_a$. Hence, provided the neuronal and synaptic variables required for computation are available in the processor, SNNs offer a path to more efficient computation if the inequality

$$C_a p\left(\frac{T}{\Delta t}\right) < C_m$$

holds. Hence, it is important to develop algorithms for SNNs that minimize $p$ and $(T/\Delta t)$ to improve computational efficiency. This requires the use of sparse binary signal encoding schemes that go beyond rate coding that is typically used in SNNs today. The following section will discuss strategies to develop general-purpose learning rules for SNNs that satisfy such constraints.

The second potential for efficiency improvement of SNNs as compared to second-generation networks arises thanks to novel memory-processor architectures based on memristive devices. While SNNs can be implemented using Si CMOS SRAM or DRAM technologies, the advent of novel nanoscale memristive devices provides opportunities for significant improvements in overall computational efficiency.

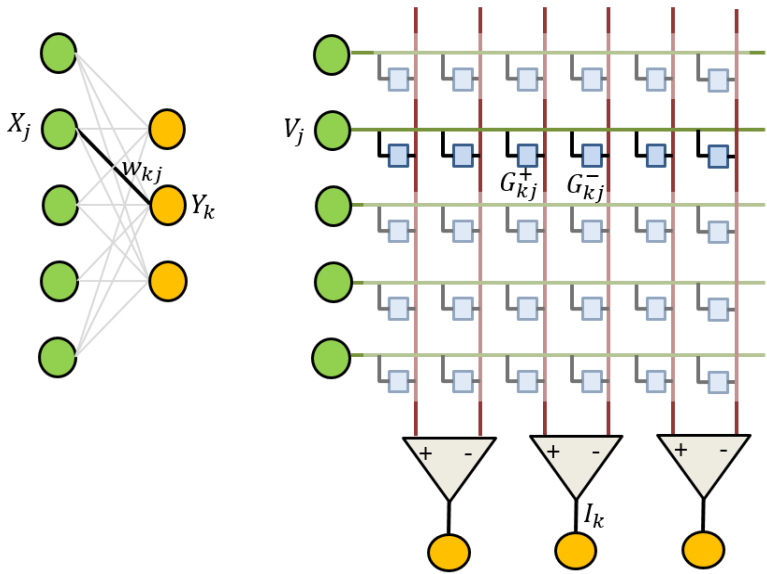

**Figure 8.** *A cross-bar array based representation of an SNN. Each synaptic weight is represented by the differential conductance of two nanoscale devices in the crossbar.*

Memristive devices can be integrated at the junctions of crossbar arrays to represent the weights of synapses, and CMOS circuits at the periphery can be designed to implement the neuronal integration and learning logic. As mentioned above, this architecture enables the computation of spike propagation operation in an efficient manner based on Kirchhoff's law as:

$$I_k = \sum_j \left(G_{kj}^+ - G_{kj}^-\right) V_j$$

In this formula, $V_j$ denotes the applied voltage pulses that are triggered when an input neuron spikes and are applied to the line connected to the $j$th input neuron, $G_{kj}^+$ and $G_{kj}^-$ are the conductances of the devices configured in a differential configuration to represent the synaptic weight, and $I_k$ is the total incoming current into the $k_{\text{th}}$ output neuron. The small form factor of the devices, coupled with the scalability of operating voltages and currents beyond what is possible with conventional CMOS, suggests that these architectures can have several orders of magnitude efficiency improvement over Silicon based implementations [68,69].

However, apart from the already mentioned non-idealities of memritive devices, crossbar arrays with more than 2048x2048 devices cannot be fabricated and operated reliably due to the resistance drop on the wires and the sneak-paths that corrupt the measurement and programming of synaptic states. One approach to mitigate these issues is to design neuro-synaptic cores with smaller crossbars and associated neuron circuits, tile these cores on a 2D array, and provide communication fabrics between the cores [70]. Such tiled neurosynaptic core-based designs are particularly amenable for realizing SNNs, as only binary spikes corresponding to intermittently active spiking neurons need to be transported between cores, as opposed to real-valued neuronal variables that are active for all the neurons in the core in the case of deep learning networks. This is the second inherent advantage that SNNs have over DNNs for computational efficiency improvement.

Overcoming the reliability challenges mentioned above is essential for building reliable systems, and would require the co-optimization of algorithms and architectures that are designed to mitigate or leverage these non-ideal behaviours for computation. Two kinds of systems can be visualized based on the application mode. Inference engines, which do not support on-chip learning, can be designed based on memristive devices integrated on crossbars, where the devices are programmed to the desired conductance states based on the weights obtained from software training. However, as memristive devices support incremental conductance changes by the application of suitable electrical programming pulses, it is also possible to design learning systems where network weight updates are implemented on-chip in an event-driven manner [82]. There are also many recent examples where these devices have been engineered to mimic the integration and fire characteristics of biological neurons [71, 72,73], potentially enabling the realization of all-memristor implementations of spiking neural networks [74]. The field is still in its infancy, and so far, has only witnessed small proof-of-concept demonstrations. We now discuss some of the approaches that have been explored towards realizing memristive based inference-only spiking networks as well as learning networks with SNNs.

4.1. **Memristive SNNs for inference**. A common approach to develop SNNs is to start with a second-generation ANN trained using traditional backpropagation-based methods, and then convert the resulting network to a spiking network in software. These solutions are based on weight-normalization schemes so that the spike rates of the neurons in the SNN are proportional to the activations of the neurons in the ANN [75, 76]. While this should in principle result in SNNs with comparable accuracies as their second-generation counterparts, some device-aware re-training would typically be necessary when the network is implemented in hardware due to the non-linearity and limited dynamic ranges of nanoscale devices.

One of the differentiating features of inference engines is that the nanoscale devices storing state variables are programmed only rarely, compared to the number of reads (potentially at every inference cycle). Since higher-energy programming cycles have a stronger effect in degrading device lifetimes compared to the lower-energy read cycles, this mode of operation can have better overall system reliability compared to that of learning systems.

In a preliminary hardware demonstration leveraging this approach, R. Midya *et al.* used memristors based on SiOxNy:Ag to implement compact oscillatory neurons whose output voltage oscillation frequency is proportional to the input current [77]. In this proof-of-concept demonstration of a 3-layer network, ANN to SNN conversion was limited to the last layer alone, but the approach could be extended to hidden layers as well.

4.2. **Memristive SNNs for unsupervised learning and adaptation.** Most hardware demonstrations of SNNs using memristive devices have focused on the unsupervised learning paradigm, where the synaptic weights are modified in an unsupervised manner according to the biologically inspired spike timing dependent plasticity (STDP) rule [78]. The rule captures the experimental observation that when a synapse experiences multiple pre-before-post pairings, the effective synaptic strength increases, and conversely, multiple post-before-pre spike pairs result in an effective decrease of synaptic conductance.

It should be noted that while other biological mechanisms may also play a key role in learning and memory formation in the brain, as have been observed experimentally[79, 80], STDP is a simple local learning rule which is especially straight-forward to implement in hardware. While it is possible to implement timing dependent plasticity rules using many-transistor CMOS circuits [81], it was experimentally demonstrated early on that memristive devices can exhibit STDP-like weight adaptation behaviours upon the application of suitable waveforms [82, 83,84]. Going beyond individual device demonstrations, IBM has also demonstrated an integrated neuromorphic core with 256x256 phase change memory synapses fabricated along with Si CMOS neuron circuits capable of on-chip learning based on a simplified model of STDP for auto-associative pattern learning tasks [85].

Boybat *et al.* used phase change memristive synapses to demonstrate temporal correlation detection through unsupervised learning based on a simplified form of STDP [86] as shown in Figure 9. In their experiment, a multi-memristive architecture was introduced, where $N$ PCM devices are used to represent one synapse, with all devices within a synapse read during spike transmission, but only one of the devices, selected through an arbitration scheme, is programmed to update the synaptic weight. Software equivalent accuracies could be obtained in the experiment with this scheme, although the individual devices are plagued by several common non-ideal effects such as programming non-linearity, read noise, and conductance drift. Note that with $N = 1$ device representing a synapse, the network accuracy was significantly lower than the software baseline; $N = 7$ devices were necessary to obtain close to ideal performance.

Spiking networks can also be used for other unsupervised learning[87] and adaptation tasks. Recently, Y. Fang *et al.* demonstrated that certain optimization problems could be solved driven by the coupled dynamics of ferroelectric field-effect transistor (FeFET) based spiking neurons [88]. While there was no synaptic weight adaptation in this approach, the optimal solution to the problem is determined by the coupled interactions between the neurons which modulate each other's membrane potentials in an event-driven manner.

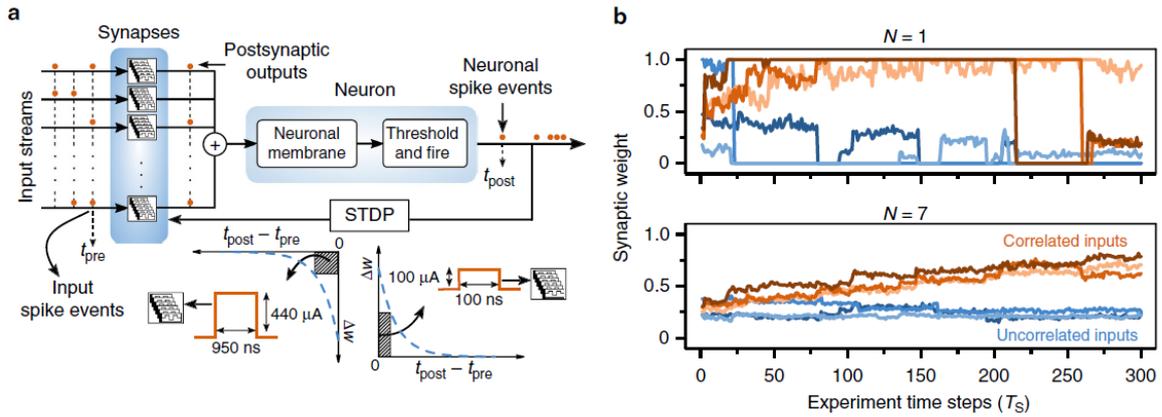

**Figure 9.** *a) Unsupervised learning demonstration using multi-memristive PCM architecture. The network consists of an integrate and fire neuron receiving inputs from 1000 multi-PCM synapses, with each synapse being excited by Poisson generated binary spike streams. 10% of the synapses receive correlated inputs, while the rest receive uncorrelated inputs. The weights evolve based on the simplified STDP rule shown. b) With N=7 PCM device per synapse, the correlated and uncorrelated synaptic weights evolve to well-separated values, while with N=1, the separation is corrupted due to programming noise. Adapted with permission [86], Copyright 2018, Nature Research.*

4.3. **Memristive SNNs for supervised learning.** Compared to the previous two approaches, implementing supervised learning in SNNs is a more challenging tasks, as the algorithm and the network must generate spikes at precise time instants based on the input excitation. As opposed to the backpropagation algorithm that is highly successful in training ANNs, supervised learning algorithms for SNNs are not well developed yet, due to the inherent difficulty in applying gradient descent methods for spiking neuron models with infinite discontinuities at the instants of spikes. Nevertheless, there have been several demonstrations of supervised learning algorithms for SNNs based on approximate forms of gradient descent for simple fully-connected networks [89,90,91].

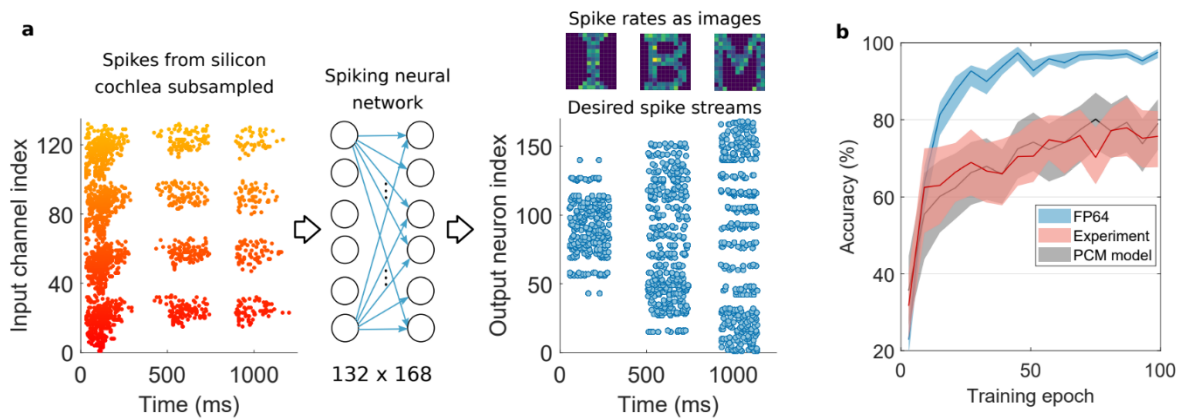

**Figure 10.** *a) SNN supervised learning experiment. A two-layer network is tasked with generating 1000ms long spike streams from the 168 neurons at the output corresponding to the images of the spoken characters. The inputs to the network are 132 spike streams representing the characters subsampled from the output of a Silicon cochlea chip. The weights are modified based on the NormAD learning rule. b) Using multi-PCM synapses, the accuracy of spike placement at the output is about 80%, compared to the FP64 accuracy of close to 98%.* [92]

Recently, Nandakumar et al., demonstrated a proof-of-concept realization of supervised learning in a 2-layer SNN implemented using nanoscale phase change memory synapses based on the Normalized Approximate Descent Algorithm [89]. In the experiment, 132 spike-streams representing spoken audio signals generated using a Silicon cochlea chip was used as the input, and the network was trained to generate 168 spike-streams whose arrival times indicate the pixel intensity corresponding to the spoken characters [92]. Compared to normal classification problems in deep networks where the accuracy depends only on the relative magnitude of the response of the output neurons, the SNN problem is harder as the network is tasked with generating close to 1000 spikes at specific time instances over a period of 1250 ms from 168 spiking neurons that are excited by 132 input spike streams. The accuracy for spike placement obtained in the experiment was about 80% compared to the software baseline accuracy of over 98%, despite using the same multi-memristive architecture described earlier. This experiment is hence illustrative of the need for developing more robust and event-driven learning algorithms for SNNs that can mitigate or even leverage the device non-idealities for designing computational systems.

4.4. **Harnessing Randomness for Learning Noise – from impairment to asset**. As discussed in the previous section, the implementation of standard deterministic learning rules, such as STDP or gradient-based schemes like NormAD [89], may be severely impaired in hardware implementations whose components are inherently noisy. In this section, we explore the idea that, if properly harnessed, native hardware randomness can be an asset for the deployment of training algorithms for SNNs [93, 94]. The gist of the argument is that randomness enables the native implementation of probabilistic models, which otherwise would require the deployment of additional, potentially costly, components. As we elaborate on next, probabilistic models have several advantages over their conventional deterministic counterparts. We focus the discussion on the problem of training, but we will also mention some of the advantages in terms of inference.

4.5. **Training deterministic SNN models.** Standard Artificial Neural Network (ANN)-based models only account for uncertainty at their inputs or outputs, while the process transforming inputs to outputs is deterministic. While limiting their expressiveness and their capacity to model structured uncertainty [95], this modelling choice does not cause a problem in the development of learning rules for ANNs. This is because deterministic ANN models define a differentiable input-output mapping as a function of the model weights, enabling the direct derivation of gradient-based learning rules through backpropagation and automatic differentiation.

Not so for SNNs. In fact, deterministic spiking neuron models such as Leaky Integrate and Fire (LIF) define non-differentiable functions of the synaptic weights: Increasing or decreasing the synaptic weights of a spiking neuron may cause the membrane potential to cross or step back from the spiking threshold, causing an abrupt change in the output. The derivative with respect to the weights is hence zero except around the firing threshold, where it is undefined. As a result, standard gradient-based learning rules cannot be directly derived for deterministic models of SNNs.

A second important issue with conventional gradient-based methods when applied to deterministic SNN models concerns the problem of credit assignment. Discrete-time deterministic SNN models can be interpreted as Recurrent Neural Networks (RNNs) with state defined by the neurons' membrane potential, input currents, and previous spiking behaviours [91]. Accordingly, the outputs and the state transition produced as a function of exogeneous inputs and state depend on the learnable synaptic weights. Therefore, a synaptic weight affects the loss function being optimized via changes that are propagated through the neurons and through time. Assigning credit for changes in the output – which is what is needed to compute the gradient – hence requires to either *backpropagate* per-output changes through neurons and time or to keep track of per-weight changes in a *forward* manner through neurons and time [96,97,91]. Both solutions come with significant drawbacks: Backpropagation requires keeping track of forward activations and flowing information backward through time, while forward methods entail the need to memorize per-weight quantities across all neurons.

Given the two challenges discussed above – non-differentiability and credit assignment – state-of-the-art training methods for SNNs based on deterministic, typically LIF, models follow various heuristic approaches. As discussed in the previous section, the most common class of methods sidesteps both challenges by carrying out an offline conversion from a pre-trained ANNs. This makes it impossible to implement online on-chip learning, and it also limits information processing to rate encoding, which encodes information in the spike frequency (e.g. see [75]). A second, popular, approach is to implement biologically inspired local synaptic update rules, such as STDP, that do not require credit assignment. The main downside of these approaches is that they do not optimize specific objective functions – sidestepping the problem of non-differentiability – and hence they are difficult to generalize to a variety of tasks and requirements. When focusing on rate encoding, it is possible to overcome to problem of non-differentiability, but not that of credit assignment, by removing non-linearities and working directly with spiking rates, for example with low-pass filtered spike trains [98,89].

In contrast to standard rate encoding, SNNs enable a novel type of information processing that computes with time, rather than merely over it as for ANNs. In order to make use of this unique capability of SNNs, it is necessary to derive learning rules that are capable of processing information encoded in the timing of the spikes and not only in their frequency. The simplest way to do this is to limit the number of spikes per neuron to one, so as to assign a continuous-valued output to each neuron. This allows the derivation of backpropagation-based rules as for ANNs, whereby the neurons' (differentiable) non-linearities capture the relationship between input and output spike timings [99].
More sophisticated methods, allowing for multiple spikes per neuron, are either based on soft non-linearity models [100] or on surrogate gradient methods [91]. The first type of approaches tackles the problem of non-differentiability by approximating the threshold activation function with a differentiable function [100]. As a result, these methods do not preserve the key feature of SNNs of processing and communicating binary spikes. The second class of techniques approximates the derivative of the threshold activation function (but not the function itself) when computing gradients [91]. Both types of methods require backward or forward propagation or the implementation of heuristic credit assignment methods such as random backpropagation [101]. As an example, SuperSpike uses forward propagation to carry out credit assignment over time coupled with random backpropagation for spatial credit assignment [90,91].

We emphasize again that the discussion above focused on the role of randomness in facilitating training. Randomness in SNNs can also be useful in the inference phase to enable Gibbs sampling-based Bayesian inference strategies [93, 102].

4.6. **Probabilistic SNN models.** Among their key advantages, probabilistic models allow the direct encoding of domain knowledge in the graph of connections among the constituent variables – a key feature of so-called expert systems – and the modelling of uncertainty [103].

They can also account for complex multi-modal distributions, unlike their deterministic counterparts [104]. Finally, stochastic models, even for ANNs, can both improve generalization, as in dropout regularization, and facilitate exploration of the training space [105].

Training of probabilistic models is generally conceptually more complex than for deterministic models due to the need to account for the exponentially large space of values that the hidden stochastic units can take. Note, however, that probabilistic models have provided the framework used to develop the first deep learning algorithms for ANNs in [106] through Boltzmann machines. Early training methods for general (undirected) models used Gibbs sampling or mean-field approximation, requiring an expensive cycling through the variables one at a time [107,108]. More modern approaches leverage advanced forms of approximate learning and inference via (Generalized) Expectation Maximization, Monte Carlo methods, and variational inference [106,104,109,110].

Probabilistic models for SNNs can be thought as direct extensions of the *belief networks* studied in [107,106,105] from static to dynamic models. As in belief networks, a neuron spikes probabilistically with a probability that increases with its membrane potential. In belief networks, the membrane potential of a neuron is an instantaneous function of the current spikes emitted by the incoming neurons in neuron's fan-in. In contrast, in an SNN, the membrane potential of a neuron evolves over time as for LIF models as a function of the past spiking behaviour of the neuron itself and of the neurons in its fan-in (see [111] for a review).

4.7. **Training probabilistic SNN models.** For the development of training rules, probabilistic SNN models have the fundamental advantage over their deterministic counterparts that the probability of the neurons' outputs is a differentiable function of the model parameters, including the synaptic weights. Many learning criteria can be formulated as the average over such distribution of a given loss or reward function. Specifically, in supervised and unsupervised learning, the learning problem can be formulated as the minimization of a loss function averaged over the joint distribution of data and of specific neurons in a read-out layer [112,111]; and in reinforcement learning, the goal is to minimize an average reward function dependent on the behaviour of the neurons in the readout layer [113]. Unlike deterministic SNN models, probabilistic SNN models hence allow naturally for the definition of differentiable learning criteria.

Once a learning criterion is determined based on the problem under study, training can be carried out via stochastic gradient-based rules. The key novel challenge in deriving such rules is the need to differentiate over the distribution of the neurons' outputs. Mathematically, with deterministic models, one needs to differentiate a training criterion of the type

$$L_d(\theta) = E_{X \sim D}[f_\theta(X)],$$

where the expectation is taken over the empirical distribution $D$ of the data and the model parameter $\theta$ directly affects the learning criterion $f_\theta(X)$ through the input-output function of the network. In contrast, with probabilistic models, the relevant learning criterion is of the type

$$L_p(\theta) = E_{X \sim D}[E_{Y \sim P_\theta}[f(X, Y)]],$$

in which $Y$ represents the random output of the neurons. Note that unlike the standard deterministic approach, the model parameters affect the learning performance through the distribution of the random output of the neurons.

Maximization of the criterion above can be in principle carried out via Expectation Maximization. In practice, the intractability of Bayesian inference of the hidden neurons entails the need for approximate solutions based on sampling methods and gradient-based techniques [104]. Computing stochastic gradients of $L_p(\theta)$ requires a double empirical expectation, one over the data distribution $D$ and one over the output distribution $P_\theta$. Estimators based on such samples can be derived by following one of a variety of principles, yielding different statistical properties in terms of, e.g., bias and variance [114].

While a number of techniques attempt to reuse the standard backpropagation algorithm, e.g., the "Straight-Through" estimator [105], an approach that is more suitable for the implementation of SNNs is obtained via the score, or log-likelihood, or REINFORCE method and variations thereof (see [104,109,110]). Accordingly, for given data and neurons' output samples the gradient with respect to a synaptic weight can be estimated through the correlation between the accrued loss function over time and the log-probability of the realized output for a given sample $(X, Y)$, i.e., (somewhat informally)

$$\nabla_\theta L_p(\theta) \approx f(X,Y) \nabla_\theta \log(P_\theta(Y)).$$

Intuitively, the higher the loss is, the more the negative gradient should push away from output distributions that generate such disadvantageous samples $Y$. Various improvements of the statistical properties of this estimator are reviewed in [114].

The REINFORCE gradient estimate $\nabla_\theta L_p(\theta)$ highlights not only the direct *differentiability* of generic learning criteria but also the fact that probabilistic learning rules solve the *credit assignment problem* by not requiring any form of backpropagation [105]. In contrast, a gradient-based rule that uses $\nabla_\theta L_p(\theta)$ only requires all nodes to receive a global feedback signal $f(X, Y)$, which may be computed by a central node [111]. The resulting learning procedure follows the standard three-factor rule from theoretical computer science, whereby the synaptic weights are modified based on pre- and post-synaptic recent spikes, which are locally available at each neuron, and on a global feedback signal [111]. Accordingly, the rule can be easily implemented in an online streaming fashion.

4.8. **Generalized probabilistic SNN models.** Apart from the advantages described above in terms of differentiability and credit assignment, probabilistic models can be directly extended with minor conceptual and algorithmic difficulties in various directions. First, it is possible to directly derive – technically, by selecting a categorical instead of a Bernoulli distribution in a Generalized Linear Model (GLM) for SNNs – training rules that allow for multi-valued spikes or inter-neuron instantaneous connections or, equivalently, Winner Take All (WTA) circuits [115, 102]. This is particularly important since data produced by some neuromorphic sensors incorporates a sign to indicate a positive or negative change [116]. Multi-valued spikes can also be used for time compression [117]. Second, various decoding rules, such as first-to-spike, can be directly optimized for, instead of having to rely on surrogate target spiking sequences [118]. Third, probabilistic models can provide an estimate of the uncertainty on the trained weights by means of Bayesian Monte Carlo methods [115].

Before describing some applications of the models and learning rules reviewed above, we mention briefly here alternative probabilistic formulations for SNNs. In the models discussed above, randomness is defined at the level of neurons' outputs. Alternative models introduce randomness at the level of synapses or thresholds [119,120].

**4.9. Examples.** Once an SNN is trained, it can be used as a sequence-to-sequence mapper in order to solve supervised, unsupervised, and reinforcement learning problems. Alternatively, with specific choices of the synaptic kernels and memory, the SNN can be used as a Gibbs sampler to carry out Bayesian inference with outputs encoded in the spiking rates [93, 102]. We now briefly discuss three applications that fall in the first category, one concerning supervised learning, one reinforcement learning, and one federated learning.

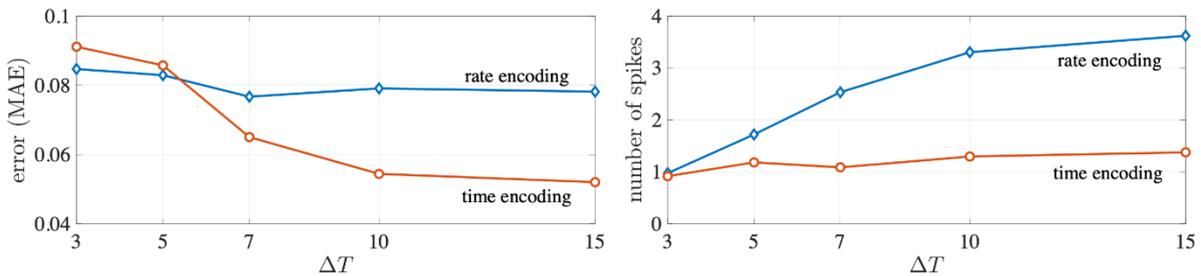

**Figure 11.** *Test error and number of spikes as a function of the time expansion parameter defining source encoding from natural signals to spikes. Reproduced with permission* [111], Copyright 2019, IEEE.

In order to first illustrate the potential of probabilistic SNNs trained to process time-encoded information, in Figure 11, we consider an online sequence prediction problem in which samples of a discrete-time source are converted into spiking signals with $\Delta T$ time instants for each sample of the input source. We consider two types of encoding, one based on standard quantization and rate encoding, and one based on the time encoding via Gaussian receptive fields. The figures, fully detailed in [111], demonstrate that time encoding can vastly outperform rate encoding both in terms of accuracy and in terms of number of spikes, which is a proxy for energy consumption.

Second, we consider a standard reinforcement learning task, in which a probabilistic SNN is used as a stochastic policy. Figure 12 compares the performance as a function of the resolution of the input grid representation for a policy directly trained with a first-to-spike decoder and one that is instead converted using state-of-the-art methods from a pre-trained ANN. The results clearly validate the intuition that directly training the stochastic policy as an SNN is more efficient than using ANN-to-SNN conversion.

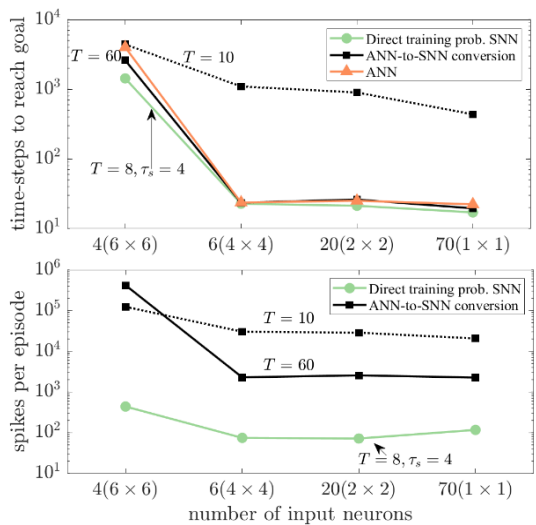

**Figure 12** *Time steps to reach goal and spikes per episode for a grid world reinforcement learning task. Reproduced with permission* [113], *Copyright 2019, IEEE.*

Finally, we consider the potential of SNN for on-mobile training via Federated Learning (FL). The approach is motivated by the fact that training on a device is limited by the amount of data available at it. Cooperative training can be carried out through FL as explored in [121], where an online FL-based learning rule is introduced for networked on-mobile probabilistic SNNs. As seen in the Figure 13 through sufficiently frequent inter-device communication, with a communication round occurring every $\tau$ iterations, the scheme demonstrates significant advantages over separate on-mobile training.

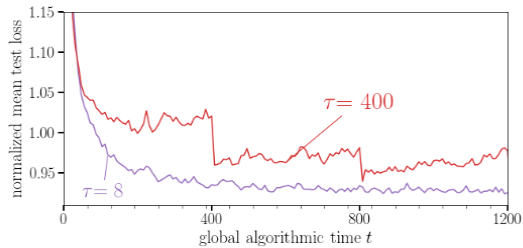

**Figure 13** *Test loss versus number of training iterations with inter-device communication taking place every $\tau$ iterations.*

4.10. **Algorithmic and hardware co-design.** To sum up the discussion in this section, spike-based learning and inference are promising facets of the neuromorphic computing paradigm. Unlike conventional machine learning models, spike-based processing "computes with time, not in time". As we have discussed, the main advantage is a potentially massive increase in power efficiency. In this section, we have presented a review of algorithmic models that leverage stochastic behavior for the implementation of SNNs. While it is true that spike-based computing can be implemented in CMOS technology, there is a great deal to be gained from compact nano-scale implementations of fundamental functional blocks – spiking neurons and adjustable synapses-- in terms of scalability and power-efficiency. Memristors are much better suited to emulate, and not merely simulate many of the sought functionalities. Moreover, the implementation of probabilistic models on current hardware platforms is made difficult by the lack of randomness sources in such systems. In contrast, the inherent randomness of switching processes in memristive devices could provide a source of randomness "for free". Research in spike-based computing is a fast-growing field. We believe that developing better-suited hardware platforms would accelerate the progress of co-designed spike-based learning and inference machines. Memristors may be the missing piece that will unlock the potential of spike-based computing.

5. **Future of neuromorphic and bio-inspired computing systems**

Taking a "big picture" view, current AI, and machine learning methods in particular, have achieved astonishing results in every field they have been applied to and have become or are becoming standard tools for nearly every type of industry one can think of. This impressive invasion was mainly propelled by deep learning which is loosely inspired by biological neural networks.

Deep learning primarily refers to learning with artificial neural networks of many layers, and fundamentally is not different to what we know about that field in the 90's. Indeed, the key algorithm underlining the success of deep learning, backpropagation, is an old story: "Learning with back-propagating errors" by Rumelhart, Hinton and Williams was published in 1986 [122]. The most commonly used neural networks are feedforward neural networks, and, convolutional neural networks used for image processing can be seen as inspired by our visual system, and both of these are not very new concepts.

Backpropagation is perhaps the most fundamental method we can think of for parameter optimisation. It is derived by differentiating an error function with respect to the learnable parameters, so in some ways it is not entirely surprising that the algorithm existed for many years. What might be somehow surprising is that we have not been able to move away much from this idea. While there has been recent progress, much of it consisted of relatively small additions and tweaks, for instance new ways to address the so called "problem of the vanishing gradient", the deterioration of the error signal as is backpropagated from the output to the input of the network. Undoubtedly, there were some fundamentally different architectures, smart techniques and novel analyses but arguably, the key factor behind such a success seems to be the vast availability of data and computational power.

In fact, recent advances of the neuroscience community are not present in the neural networks. We do not want to argue that this, per se, is either good or bad, or to suggest that the next super-algorithms will be copying nature. We only want to underline that though inspired only, artificial neural networks had their basis on neuroscience concepts and that there are many phenomena that have, perhaps, not been sufficiently explored within an AI context. For instance, biological neural networks have different learning rules for positive and negative connections, connections change in multiple time scales and show reversable dynamic behavior (known as short-term plasticity), and the brain itself has a structure where specific areas play different roles, just to name a few.

Instead, our progress was mainly based on hardware improvements that made this success possible by allowing long training phases; an amount of training unrealistic for any human. While it is true that human intelligence also develops over years and that human learning involves many trials, for comparison AlphaGoZero, which surpass human performance in the game of Go, was trained over 4.9 Million games[123]. To match this number of games would require a human that lives for 90 years to complete one Go game every 10 minutes from the moment they are born. This realization tells us two things: (1) our machines do not learn the same way that humans do, and even if we think our methods as bioinspired, we likely still miss some key ingredients and (2) executing that many games certainly require considerable computational power and energy consumption.

As a consequence, training algorithms often require a high energy footprint due to excessive training times and hyper parameter tuning involved. Hyper-parameters are parameters of the system that are not (usually) adapted via the learning method itself, one such example is the learning rate, which indicate how fast the network should update its "knowledge". Before rushing to say that a high learning rate is obviously desirable, such a learning rate could lead to oscillations as, for instance, optimal solutions could be overlooked, or it could lead to forgetting previously obtained knowledge. Setting the learning rate right is not always trivial. In fact, the tuning of hyper-parameters was what originally made the machine learning community to turn away from artificial neural networks, and it was the performance of deep learning that brought the focus back. One may then wonder, at the end of the day how much energy inefficient could deep learning systems be? The reply is perhaps surprising: estimated carbon emissions for training standard natural language processing models is approximately five times higher than running a car for a lifetime[124]. This realization suggests there is an urgent need to improve on both current hardware and learning models.

Given such energy concerns, systems based on low-power memristive devices are a highly promising alternative [125,126]. Besides having a low carbon footprint, there is numerus work demonstrating devices that mimic neurons, synapses, and plasticity phenomena. Often such approaches work well for offline training. However, some of these attempts, particularly where plasticity is involved, are opportunistic (including own work) and how scaling to larger networks could happen is not always obvious. Faithfully reproducing the brain functionality, when neuroscience has already so many open questions is challenging for any technology. Moreover, using technologies that potentially allow less possibilities for engineering in comparison to traditional methods (such as CMOS) might well be mission impossible. How far can we go by reconstructing neuron by neuron and synapse by synapse in terms of scalability remains unclear. A more promising way might be to achieve a deeper understanding of the physics of the relevant materials and based on this understanding co-develop the technology and the required learning methods for achieving Artificial Intelligence.

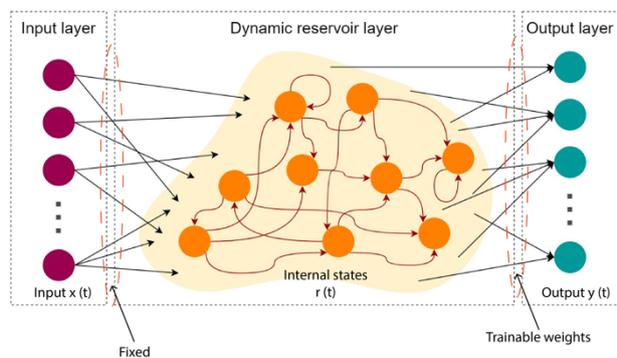

**Figure 14** Reservoir Computing maps inputs x(t) to higher-dimensional space, defined by the reservoir states r (t). Only weights connecting reservoir states r (t) and output y (t) need to be trained.

In the meantime, in parallel, we can immediately explore simple bio-inspired approaches that harness the dynamics of the material and could be proven useful for particular sets of problems. Here we present one such example which stems from the area of reservoir computing, an idea invented separately by Herbert Jeager for the machine learning community[127], under the name of echo state networks, and by Wolfgang Mass[128] for the computational neuroscience community, under the name of liquid state machines. We strongly suspect that both these methods were very much motivated by the difficulty of training recurrent networks with a generalization of backpropagation known as backpropagation through time. While feedforward networks can perform many tasks successfully, recurrences are required for memory and, moreover, the brain is clearly not only feedforward. If recurrences exist and are required, there must be a way to efficiently train such structures. As a side note, it is very difficult to imagine how a biological neural network would be able to implement backpropagation through time, and for this alternative approaches have recently made their appearance[129].

Reservoir computing methods came up with a workaround to the problem of training recurrent networks: they do not train them but instead harness their properties. Common in the approaches of echo state networks and liquid state machines is the idea of using a randomly recurrent network with fixed connectivity, hence no need to resort to backpropagation through time. This recurrent network is called a reservoir. It provides memory and at the same time transforms the input data to a spatiotemporal representation of higher dimensionality. This enhanced representation can be used as an input to single-layer perceptrons, that are trained with a very simple learning method, so the only learnable parameters are the feedforward weights between the reservoir neurons and the output neurons. The key difference between the echo state networks and liquid state machines, is that the first approach uses recurrent artificial neuron dynamics while the second uses recurrent spiking neural networks, reflecting the mindset in their corresponding communities. The main principle of reservoir computing is shown in Figure 14. The input x (t) is projected into the higher dimensional feature space r (t) by using the dynamical reservoir system. Only the weights connecting the internal states r (t) with the output y (t) need to be trained, while the rest of the system is fixed. The advantage of this approach is that it only requires a simple training method, while the ability to process complex and temporal data is retained.

Indeed, it might be surprising how much randomness can do from the point of computation: a random network can enrich data representations sufficiently so that a linear method can separate the data into the desirable classes. This approach is conceptually similar to the well-known method of support vector machines, which uses kernels to augment the dimensionality of the data, so that again only a simple linear method is sufficient to achieve data classification. In fact, a link between the purely statistical technique of support vector machines and the bioinspired technique of reservoir computing has been formally built [130]. We can perhaps think of this link as a demonstration that biological inspiration and purely mathematical methodology might also solve problems in a similar manner.

We claim that reservoir computing would benefit from appropriate hardware. When simulating, the convergence of the recurrent network requires time, because the continuous system will be discretized and sequentially run on the CPU. If instead we replace the reservoir with an appropriate material, this step could become both fast and energy efficient: the material could compute effortlessly using its physical properties. Reservoirs do not need overengineering, since no specific structure is required; we only need to produce dynamics that are complex enough but not chaotic. In fact, there has already been work exploiting memristors in this direction[131].

Could ideas from biology still add value to existing methods? A recent augmentation of the echo state networks[132], inspired by the fruitfly brain, explores the concept of sparseness in order to improve learning performance of reservoirs. In brains, contrary to the typical artificial neural networks, only few neurons fire at a time, a fact that has been linked to memory capacity. Neuronal thresholds, appropriately initialized and updated with a slower time constant than that of the feedforward learnable weights can modulate sparseness and lead to better performance in comparison to the non-sparse reservoir, but also in comparison to state-of-the-art methods in a set of benchmark problems. Due to the sparseness leading to task-specific neurons, this bio-inspired technique can alleviate the problem of catastrophic forgetting. Machine learning methods often suffer from the fact that once they learn a new task they have forgotten the previous one. Since in the space reservoir network a new task will likely recruit previously unused neurons, leaning a new skill does not completely override those previously learned. This simple method competes and surpasses more complicated methods that are built specifically to address catastrophic forgetting. Most importantly, the formulation of the specific rule allows for completely replacing the network dynamics with any other dynamics, including material dynamics, that are suitable for the purpose (i.e. highly nonlinear and not chaotic). Perhaps there are more such lessons to be learned from biology.

So, what can be done right now? To us it is clear that a better understanding of the physics behind memristive devices is key for the progress of the field[133]. A deeper understanding will allow us to harness the properties of the system for brain-like computation rather trying to fabricate some arbitrary brain behavior that may or may not be important in the context of a specific application, or worse may not scale up. Instead of thinking at the level of mimicking neurons and synapses, we can instead take inspiration from the biological systems, consider the dynamics required for neuronal processing and use the material physics to reproduce them.

6. **Conclusion**

Memristor technologies are still to realise their full potential that has been promoted over the last 15 years. Although predominantly seen as candidates to replace or augment our current digital memory technologies, the impact of memristor technologies on the broader fields of artificial intelligence and cognitive computing platforms are likely to be even more significant. As discussed in this progress report, the versatility of memristor technologies has resulted in their use across a range of applications: from in-memory computing, deep learning accelerators, spiking neural networks, to more futuristic bio-inspired computing paradigms. These approaches should not be seen as solutions to the same problem, nor as technologies that are in direct competition among themselves or with current, very successful, CMOS systems. Additionally, it is crucial to recognise that many of the discussed research areas are still at the very beginning of their development. Of these, more mature approaches will likely produce industrially relevant solutions sooner. For example, greater power-efficiency is an essential utility and a pressing issue that many engineers are trying to address. In-memory computing and deep learning accelerators based on memristors represent an attractive proposition for extreme power-efficiency.

There is also significant scope for more fundamental work. Development of new generations of bio-inspired algorithms would further boost advancements in hardware systems and platforms. The challenge and opportunity lie in the interdisciplinary nature of the research and the necessity to understand distinct methodologies and approaches. We believe that the community will benefit from the next generation of researchers being well educated across different traditional disciplines. For example, there is an undeniable link between the fields of computer science, more specifically machine learning, and computational neuroscience. The two disciplines could co-exist separately and act independently with distinct goals; however, there are great benefits to be gained from a more holistic approach. A strong case for closer collaboration has been made recently [134]. Collaborations should be expanded to include researchers in solid-state physics, materials science, nanoelectronics, circuit/architecture design and information theory. Memristors show great promise to be a fabric for producing brain-inspired building blocks [135], and this progress report showcases different types of memristor-based applications. Memristor technologies are versatile enough to provide the perfect platform for different disciplines to strive together in pushing the frontiers of our current technologies in the most fundamental way.

**Acknowledgements**
A.M. acknowledges funding and support from the Royal Academy of Engineering under the Research Fellowship scheme. A.S. acknowledges funding from the European Research Council (ERC) under the European Union's Horizon 2020 research and innovation programme (grant agreement number 682675). B.R. acknowledges partial support from IBM and Cisco. O. S. acknowledges funding from the European Research Council (ERC) under the European Union Horizon 2020 research and innovation program (grant agreement 725731). E.V. would like to acknowledge a Google Faculty Research Award (2017). AJK acknowledges funding from the Engineering and Physical Sciences Research Council (EPSRC).